\documentclass[12pt]{article}
\usepackage{epsf}
\usepackage{amsmath}
\usepackage{graphics}
\usepackage{cite}

\renewcommand{\thefootnote}{\#\arabic{footnote}}
\begin{document}

\newcommand{\gtrsim}{ \mathop{}_{\textstyle \sim}^{\textstyle >} }
\newcommand{\lesssim}{ \mathop{}_{\textstyle \sim}^{\textstyle <} }

\newcommand{\rem}[1]{{\bf #1}}

\renewcommand{\thefootnote}{\fnsymbol{footnote}}
\setcounter{footnote}{0}
\begin{titlepage}

\def\thefootnote{\fnsymbol{footnote}}

\begin{center}
\hfill hep-ph/0701034v3

\vskip .5in
\bigskip
\bigskip
{\Large \bf Group Theoretic Bases for Tribimaximal Mixing}

\vskip .45in

{\bf Paul D. Carr and Paul H. Frampton} 

{\em Department of Physics and Astronomy, University of North Carolina,}

{\em Chapel Hill, NC 27599-3255. }

\end{center}

\vskip .4in

\begin{abstract}
Present data on neutrino masses and mixing favor
the highly symmetric tribimaximal neutrino
mixing matrix which suggests an underlying 
flavor symmetry. A systematic study of non-abelian finite groups
of order $g \leq 31$ reveals that tribimaximal mixing
can be derived not only from the well known flavor symmetry 
$T \equiv A_4$, the tetrahedral group,
but also by using the alternative flavor symmetry
X(24) $\equiv SL_2(F_3) \equiv Q_4 \tilde{\times} Z_3$.
X(24) does not contain the
tetrahedral group as a subgroup, and has the advantage
over it as a flavor symmetry that it can not only
underwrite bitrimaximal mixing for neutrinos, equally
as well, but also provide a first step to 
understanding the quark mass hierarchy.

\end{abstract}

\end{titlepage}

\renewcommand{\thepage}{\arabic{page}}
\setcounter{page}{1}
\renewcommand{\thefootnote}{\#\arabic{footnote}}

\newpage

\bigskip

Progress in our knowledge of the three neutrino masses and mixings
has been remarkable since SuperKamiokande found the first convincing
evidence of non zero neutrino mass in 1998\cite{superKamiokande}, and
the Sudbury Neutrino Observatory (SNO) exceeded all expectations by
abruptly solving the solar neutrino puzzle in 2001\cite{SNO}
thereby resolving the 35-year old conundrum set up by the 
persistent, and correct, experiments by Davis\cite{Davis}.
It is probably fair to say that previously 
the majority of colleagues believed the data of Davis were explicable 
by suspected inaccuracies of the Standard Solar Model (SSM), but as we now know the
SSM is a description of our Sun which is accurate to much better
than a factor three, actually to within ten per cent\cite{SNO}.
It is fair to say that our present knowledge of neutrino flavor is at least
comparable to that of quark flavor despite the fact that the theory 
for quark
flavor goes back to the 1963 article by Cabibbo\cite{Cabibbo}
(pre saged by a footnote in the 1960 paper by Gell-Mann and L\'{e}vy\cite{GMLevy})
and the paper by Gatto {\it et al.} in 1968\cite{Gatto}.  
No complete understanding 
of the quark masses and mixings has subsequently emerged and 
the prediction of CP violation in \cite{Kobayashi} provides no 
insight into its magnitude. 

We shall consider only three 
left-handed neutrinos at first, so avoiding any encounter 
with the see-saw mechanism\cite{seesaw}. The Majorana mass matrix
${\cal M}$ is a $3 \times 3$ unitary symmetric matrix and without
CP violation has six real parameters.
Let write the diagonal form as ${\bf M} = {\rm diag} (m_1, m_2, m_3)$,
related to the flavor basis ${\cal M}$ by ${\bf M} = U^{T} {\cal M} U$
where $U$ is orthogonal. It is the form of ${\cal M} = U{\bf M}U^{T}$
and $U$ which are the targets of lepton flavor physics. 
One technique for analysis of ${\cal M}$ is to assume texture 
zeros\cite{textures,marfatia, FGY} in ${\cal M}$ and this
gives rise to relationships between the mass eigenvalues $m_i$
and the mixing angles $\theta_{ij}$. For example, it
was shown in \cite{marfatia} that ${\cal M}$ cannot have
as many as three texture zeros out of a possible six but can
have two.
A quite different interesting philosophy is that neutrino masses may arise from
breaking of lorentz invariance\cite{Glashow}. Clearly, a wide range
of approaches is being aimed at the problem.

In the present study we focus on a symmetric texture for ${\cal M}$ 
with only four independent parameters, of the form

\begin{equation}
{\cal M} = \left(
\begin{array}{ccc}
A & B & B \\
B & C & D \\
B & D & C 
\end{array}
\right)
\label{symmetric}
\end{equation}

\noindent The ${\cal M}$ can be reached from a diagonal ${\bf M}$ by the orthogonal
transformation
\begin{equation}
U = \left(
\begin{array}{ccc}
{\rm cos} \theta_{12} & {\rm sin} \theta_{12} & 0 \\
- {\rm sin} \theta_{12}/\sqrt{2} & {\rm cos} \theta_{12} / \sqrt{2} &
- 1/\sqrt{2} \\ 
- {\rm sin} \theta_{12}/\sqrt{2} & {\rm cos} \theta_{12} / \sqrt{2} &
1/\sqrt{2} 
\end{array}
\right)
\label{theta12}
\end{equation}
where one commits to a relationship between $\theta_{12}$
and the four parameters in Eq.(\ref{symmetric}), namely

\begin{equation}
{\rm tan} 2 \theta_{12} = 2 \sqrt{2B (A-C-D)^{-1}}
\end{equation}

\noindent Written in the standard PMNS form\cite{PMNS}
\begin{equation}
U =
\left(
\begin{array}{ccc}
1 & 0 & 0 \\
0 & {\rm cos} \theta_{23} & {\rm sin} \theta_{23} \\
0 & - {\rm sin} \theta_{23} & {\rm cos} \theta_{23} \\
\end{array}
\right)
\left(
\begin{array}{ccc}
{\rm cos} \theta_{13}  & 0 & {\rm sin} \theta_{13} e^{i \delta} \\
0 & 1  & 0 \\
- {\rm sin} \theta_{13} e^{- 1 \delta}  & 0  & {\rm cos} \theta_{13} \\
\end{array}
\right)
\left(
\begin{array}{ccc}
{\rm cos} \theta_{12}  & {\rm sin} \theta_{12}  & 0 \\
- {\rm sin} \theta_{12}  & {\rm cos} \theta_{12}  & 0 \\
0 & 0  & 1 
\end{array}
\right)
\end{equation}
this ansatz requires that $\theta_{23} = \pi/4$ and $\theta_{13} = 0$,
both of which are consistent with present data. These
values of maximal $\theta_{23}$ and vanishing $\theta_{13}$ are
presumably only approximate but departures therefrom, if
they show up in future experiments, could be accommodated by 
higher order corrections.

\noindent To arrive at tribimaximal 
mixing\cite{Cabibbo2,Wolfenstein,Ma,Ma2,HPS,Altarelli}, 
one more parameter $\theta_{12}$
in Eq. (\ref{theta12}) is assigned such that the entries of the
second column are equal, {\it i.e.} ${\rm sin} \theta_{12}
= {\rm cos} \theta_{12}/\sqrt{2}$ which implies that
${\rm tan}^2 \theta_{12} = 1/2$. Experimentally $\theta_{12}$
is non-zero and over $5\sigma$ from a maximal $\pi/4$.
The present value\cite{expt} is ${\rm tan}^2 \theta_{12} = 0.452^{+0.088}_{-0.070}$,
so the tribimaximal value is within
the allowed range. With this identification Eq.(\ref{theta12}) becomes\cite{HPS}

\begin{equation}
U_{HPS} = \left(
\begin{array}{ccc}
\sqrt{2/3} & \sqrt{1/3} & 0 \\
- \sqrt{1/6} & \sqrt{1/3} & - 1/\sqrt{2} \\ 
- \sqrt{1/6} & \sqrt{1/3} & 1/\sqrt{2} 
\end{array}
\right)
\label{HPS}
\end{equation}

\noindent This ensures that the three mixing angles $\theta_{ij}$ are consistent
with present data, although more accurate experiments may require
corrections to $U_{HPS}$. 

The data allow the normal hierarchy which occurs most often from models
with $|m_3| \gg |m_{2,1}|$. In the normal hierarchy one expects
$|m_3| \sim \sqrt{|\Delta_{23}|} \sim 0.05$ eV, $|m_2| \sim
\sqrt{|\Delta_{12}|} \sim 0.009$ eV and $|m_1|$ essentially zero, as is
the prediction for the eigenvalues in the FGY model\cite{FGY}.
The data also allow for an inverted hierarchy with
$|m_1| \sim |m_2| \gg |m_3|$. 
A third possible pattern is the degenerate case $|m_1| \sim |m_2| \sim |m_3|
\gg |(m_3 - m_2)|$.
The tribimaximal mixing, $U_{HPS}$  of Eq.(\ref{HPS}), can accommodate all
three of these neutrino mass patterns and so makes no prediction
in that regard.

The success of $U_{HPS}$ tribimaximal neutrino mixing has precipitated
many studies of its group theoretic basis\cite{Ma,Ma2,Altarelli} and
the tetrahedral group $A_4$ has emerged.
was prompted by earlier work of one of the present authors (PHF)
with Kephart in systematically studying {\it all} non-abelian
finite groups of order $g \leq 31$ both for a quark flavor group
\cite{Kephart1} and for orbifold compactification in string theory
\cite{Kephart2}. Our 
question is whether or not $A_4$ is singled out from these as the neutrino
flavor symmetry?

\begin{center}

{\bf Character Table of $T$}

$\omega = {\rm exp}(2 \pi i / 3)$

\bigskip
\bigskip

\begin{tabular}{||c||c|c|c|c||}
\hline\hline
 & $1_1$  & $1_2$ & $1_3$ & 3 \\
\hline\hline
$C_1$ & 1 & 1 & 1 & 3 \\
\hline
$C_2$ & 1 & 1 & 1 & -1 \\
\hline
$C_3$ & 1 & $\omega$ & $\omega^2$ & 0 \\
\hline
$C_4$ & 1 & $\omega^2$ & $\omega$ & 0 \\
\hline\hline
\end{tabular}

\bigskip
\bigskip
\bigskip
\bigskip

{\bf Kronecker Products for Irreducible Representations of $T$}

\bigskip
\bigskip

\begin{tabular}{||c||c|c|c|c||}
\hline\hline
 & $1_1$  & $1_2$ & $1_3$ & 3 \\
\hline\hline
$1_1$ & $1_1$ & $1_2$ & $1_3$ & $3$ \\
\hline
$1_2$ & $1_2$ & $1_3$ & $1_1$ & $3$ \\
\hline
$1_3$ & $1_3$ & $1_1$ & $1_2$ & $3$ \\
\hline
$3$ & $3$ & $3$ & $3$ & $1_1+1_2+1_3+3+3$ \\
\hline\hline
\end{tabular}

\end{center}

\bigskip

The Kronecker products for irreducible representations
for all the fourty-five non-abelian finite groups with order $g\leq31$ are
explicitly tabulated in the Appendix
of \cite{Kephart2}, where the presentation is adapted to a style 
aimed at model builders in theoretical physics rather
than at mathematicians as in \cite{ThomasWood}.

Study of \cite{Kephart2} shows that a promising flavor group is
X(24) $\equiv SL_2(F_3) \equiv Z_3 \times Q$. The Kronecker products are
identical to those of $T \equiv A_4$ if the doublet representations are omitted
and so the group X(24) can reproduce successes of $A_4$ model building.
The use of X(24) as a flavor group first appeared
in \cite{Kephart1} and then analysed in more details in \cite{Lebed}.

X(24) has an advantage over $T$ in extension to the quark sector
because the doublets of X(24), absent in $T$, allow the implementation
of a $(2+1)$ structure to the three quark families, thus permitting the
third heavy family to be treated differently as espoused
in \cite{Kephart3,Kephart1,Conference}

\begin{center}

{\bf Character Table of X(24)}

$\omega = {\rm exp}(2 \pi i / 6)$

\begin{tabular}{||c||c|c|c|c|c|c|c||}
\hline\hline
 & $1_1$  & $1_2$ & $1_3$ & $2_1$ & $2_2$ & $2_3$ & $3$ \\
\hline\hline
$C_1$ & $1$ & $1$ & $1$ & $2$ & $2$ & $2$ & $3$ \\  
\hline
$C_2$ & $1$ & $1$ & $1$ & $- 2$ & $- 2$ & $- 2$ & $3$ \\  
\hline
$C_3$ & $1$ & $\omega^2$ & $\omega^4$ & $- 1$ & $\omega^5$ & $\omega$ & $0$ \\
\hline
$C_4$ & $1$ & $\omega^4$ & $\omega^2$ & $- 1$ & $\omega$ & $\omega^5$ & $0$ \\
\hline
$C_5$ & $1$ & $1$ & $1$ & $0$ & $0$ & $0$ & $- 1$  \\
\hline
$C_6$ & $1$ & $\omega^2$ & $\omega^4$ & $- 1$ & $\omega^2$ & $\omega^4$ & $0$ \\
\hline
$C_7$ & $1$ & $\omega^4$ & $\omega^2$ & $1$ & $\omega^4$ & $\omega^2$ & $0$ \\
\hline\hline
\end{tabular}

\bigskip

\newpage

{\bf Kronecker Products for Irreducible Representations of X(24)}

\bigskip
\bigskip

\begin{tabular}{||c||c|c|c|c|c|c|c||}
\hline\hline
 & $1_1$  & $1_2$ & $1_3$ & $2_1$ & $2_2$ & $2_3$ & $3$ \\
\hline\hline
$1_1$ & $1_1$ & $1_2$ & $1_3$ & $2_1$ & $2_2$ & $2_3$ & $3$  \\
\hline
$1_2$ & $1_2$ & $1_3$ & $1_1$ & $2_2$ & $2_3$ & $2_1$ & $3$  \\
\hline
$1_3$ & $1_3$ & $1_1$ & $1_2$ & $2_3$ & $2_1$ & $2_2$ & $3$  \\
\hline
$2_1$ & $2_1$ & $2_2$ & $2_3$ & $1_1 + 3$ & $1_2 + 3$ & $1_3 + 3$ & $2_1 + 2_2 + 2_3$ \\
\hline
$2_2$ & $2_2$ & $2_3$ & $2_1$ & $1_2 + 3$ & $1_3 + 3$ & $1_1 + 3$ & $2_1 + 2_2 + 2_3$ \\
\hline
$2_3$ & $2_3$ & $2_1$ & $2_2$ & $1_3 + 3$ & $1_1 + 3$ & $1_2 + 3$ & $2_1 + 2_2 + 2_3$ \\
\hline
$3$ & $3$ & $3$ & $3$ & $2_1 + 2_2 + 2_3$ & $2_1 + 2_2 + 2_3$ & $2_1 + 2_2 + 2_3$ &
$1_1 + 1_2 + 1_3 + 3 + 3$ \\
\hline\hline
\end{tabular}

\bigskip
\end{center}

It is important to remark that X(24) does not contain $T$
as a subgroup\cite{ThomasWood} so our discussion about quark masses
does not merely extend $T$, but replaces it.

\bigskip

The philosophy used for X(24) is reminiscent of much earlier
work in \cite{Conference,Kephart3} where the dicyclic
group $Q_6$ was used with doublets and singlets for the
(1st, 2nd) and (3rd) families to transform as $({\bf 2} + {\bf 1})$
respectively. On the other hand, $Q_6$ is not suited
for tribimaximal neutrino mixing because like all dicyclic groups $Q_{2n}$
it has no triplet representation. Recall that when the
work on $Q_6$ was done, experiments had not established neutrino mixing for
the reason explained in our first paragraph. 

For the quark sector, the X(24) assignments are most naturally chosen
using the set of model building steps for a quark flavor
group $G$ introduced in \cite{Kephart1}. The main purpose is to understand 
why the third family of quarks and leptons is heavy,
and especially why the top quark is {\bf very} heavy. The steps are:

(A) The t quark mass (and {\it only } the t ) transforms as a {\bf 1} of G.

(B) The b and $\tau$ masses appear as G is broken to $G^{'}$.

(C) After stage (B) first the c mass ($G' \rightarrow G''$), then the
s and $\mu$ masses ($G^{''} \rightarrow G^{'''}$) are generated.
At stage (C) the u, d and e remain massless.

(D) No additional quarks and a minimal number of leptons be introduced beyond
the usual three-family standard model.

\noindent We start by satisfying (A) through (D). 
We therefore assign the quarks as follows, with the charged leptons like down-type
quarks.

$$\begin{array}{cc}

\left. \begin{array}{c} \left( \begin{array}{c} t \\ b \end{array} \right)_{L}
{}~~~ 1\\
\left. \begin{array}{c} \left( \begin{array}{c} c \\ s \end{array} \right)_{L}
\\
\left( \begin{array}{c} u \\ d \end{array} \right)_{L}  \end{array} \right\}
2_1 \end{array} \right.&
\left. \begin{array}{c} t_{R}~~~ 1 \\ c_{R} ~~~ 1 \\ u_{R} ~~~ 1 \\
\left.\begin{array}{c}
 b_{R} \\ s_{R} \\ d_{R} \end{array} \right\} 3 \end{array} \right.
\end{array}$$  
\noindent whereupon the mass matrices are:
$$U = \left( \begin{tabular}{c|c}
$<2_1>$ & $ <1> $ \\  \hline
$<2_1>  $ & $ <1>   $
\end{tabular} \right)$$ \\
\noindent and
$$D = L = \left( \begin{tabular}{c|c}
$<2_1 + 2_2 + 2_3>$ & $ <3> $ \\  \hline
$<2_1 + 2_2 + 2_3>$ & $ <3> $
\end{tabular} \right)$$\\
\noindent To implement the hierarchy complying with rules (A) to (D) above requires:

(A) A VEV to a $SU(2)_L$ doublet which is a singlet of X(24) gives a
heavy mass to t without breaking X(24).

(B) A VEV to a {\bf 3} of X(24) gives mass to b and $\tau$.

(C) and (D)  The c quark acquires mass radiatively through
a VEV of $(1 , 2_1 )$ via the diagram of Fig. (1) below. 
The s and $\mu$ acquire mass
at tree level
through $2_1$ or $2_2$ VEVs, breaking $G^{'}$. The u, d and e are
still massless.

\begin{flushleft}
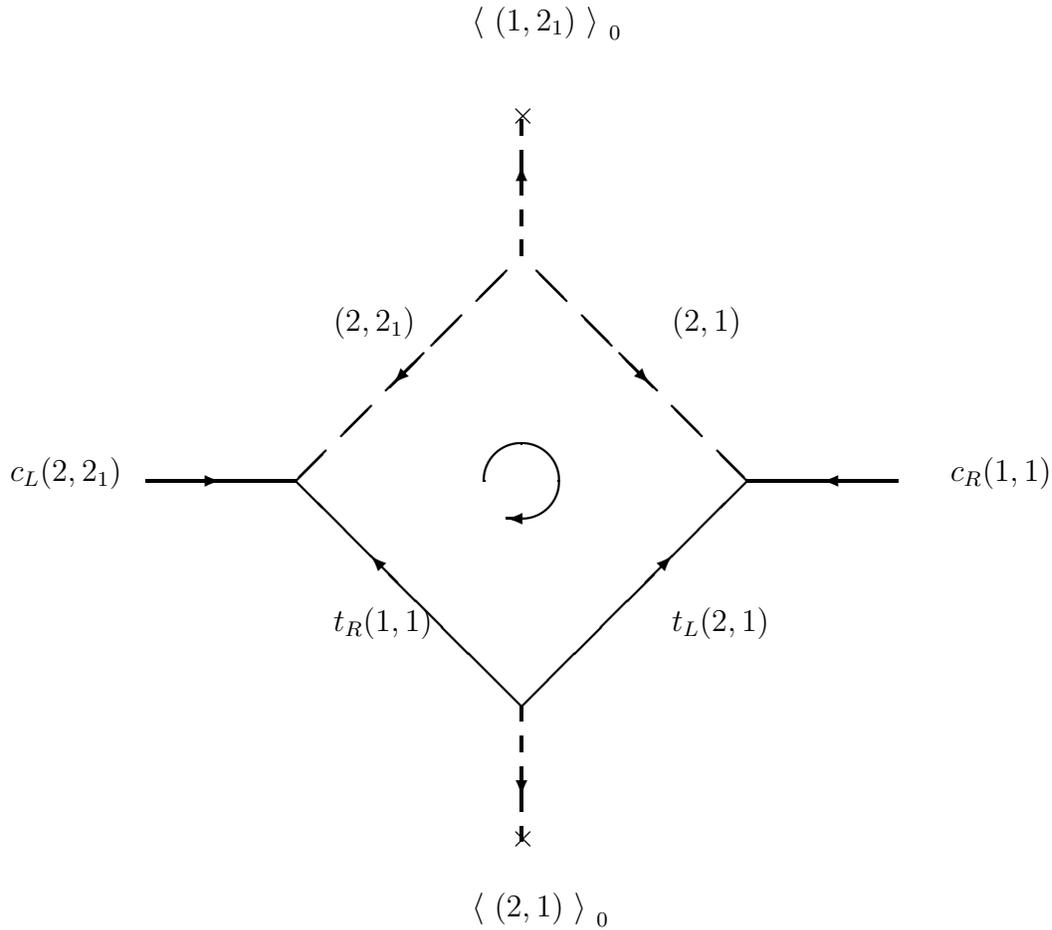
\begin{figure}[h]

\vspace*{2.0cm}

\setlength{\unitlength}{1.0cm}

\begin{picture}(15,12)

\thicklines

\put(2,6){\thicklines\vector(1,0){1}}
\put(3,6){\line(1,0){1}}

\multiput(4,6)(0.6,0.6){5}{\line(1,1){0.4}}

\put(7,3){\vector(-1,1){2}}
\put(5,5){\line(-1,1){1}}

\multiput(7,9)(0,0.4){5}{\line(0,1){0.2}}

\put(12,6){\vector(-1,0){1}}
\put(11,6){\line(-1,0){1}}

\multiput(10,6)(-0.6,0.6){5}{\line(-1,1){0.4}}

\multiput(7,3)(0,-0.4){5}{\line(0,-1){0.2}}

\put(7,3){\vector(1,1){2}}
\put(9,5){\line(1,1){1}}

\put(8.3,7.7){\vector(1,-1){0.4}}

\put(5.7,7.7){\vector(-1,-1){0.4}}

\put(7,9.8){\vector(0,1){0.4}}

\put(7,2.2){\vector(0,-1){0.4}}

\put(6.854,1.127){${\bf \times}$}

\put(6.854,10.75){$\times$}

\put(6.2,12){
$ \left \langle \ {(1,2_1)} \ \right \rangle_{\ 0} $}

\put(6.2,0.2){
$ \left \langle \ {(2,1)} \ \right \rangle_{\ 0} $}

\put(0.2,6){$ c_L (2,2_1) $}

\put(12.7,6){$ {c_R (1,1)} $}

\put(7,6){\oval(1,1)[t]}
\put(7,6){\oval(1,1)[br]}
\put(7,5.5){\vector(-1,0){0.2}}

\put(9,8){$(2,1)  $}
\put(9,4){$ t_L (2,1) $}

\put(4.5,8){$ (2,2_1) $}
\put(4.5,4){$ t_R (1,1)$}

\end{picture}

\caption{One loop diagram contributing to the charm quark mass.}

\label{Fig. 1}

\end{figure}

\end{flushleft}

\bigskip
\bigskip

In summary, while $T \equiv A_4$ is one candidate for a lepton flavor group which
gives rise naturally to tribimaximal mixing it is not 
unique among the non abelian
finite groups in this regard. 
The choice X(24) $\equiv SL_2(F_3) \equiv Z_3 \times Q$ satisfies the requirement
equally well, and because it has doublet representations
can readily accommodate the quark mass spectrum, particularly the
anomalously heavy third family.

As a flavor group to accommodate both quark and lepton masses X(24) 
emerges as a leading candidate and so a non-gravitational grand unified 
theory with aymptotic high energy symmetry

\begin{equation}
G_{GUT} \times X(24) 
\label{gut}
\end{equation}

\noindent where $G_{GUT}$ is the gauge group
of a grand unified theory is strongly suggested. 

In the present Letter we have shown how both the neutrino
mixing angles and the quark masses are naturally fitted by the choice
of flavor symmetry X(24); Nature chooses the triplet representations
of X(24) for neutrinos and both
doublet and triplet representations for
quarks and charged leptons.
The complementary goal of understanding the
neutrino masses and the quark mixing angles undoubtedly
requires dynamics associated with $G_{GUT}$
in the overall symmetry (\ref{gut}).
Time will tell whether (\ref{gut}) is the best selection of asymptotic symmetry
but we believe the present article encourages such a choice.

\bigskip
\bigskip

\begin{center}

{\bf Acknowledgements}

\end{center}

\bigskip
This work was supported in part by the
U.S. Department of Energy under Grant No. DE-FG02-06ER41418.


\bigskip
\bigskip

\begin{thebibliography}{100}
\bibitem{superKamiokande}
Y. Fukuda {\it et al.} (Super-Kamiokande Collaboration)
Phys. Rev. Lett. {\bf 81,} 1562 (1998). {\tt hep-ex/9807003}.
\bibitem{SNO}
Q.R. Ahmad {\it et al.} (SNO Collaboration) Phys. Rev. Lett. {\bf 87,} 071301 (2001).
{\tt nucl-ex/0106015}.
\bibitem{Davis}
B.T. Cleveland {\it et al.} Astrophys. J. {\bf 496,} 505 (1998).
\bibitem{Cabibbo}
N. Cabibbo, Phys. Rev. Lett. {\bf 10,} 531 (1963).
\bibitem{GMLevy}
M. Gell-Mann and M. Levy, Nuovo Cim. {\bf 16,} 705 (1960).
\bibitem{Gatto}
R. Gatto, G. Sartori and M. Tonin, Phys. Lett. {\bf B28,} 128 (1968).
\bibitem{Kobayashi}
M. Kobayashi and T. Maskawa, Prog. Theor. Phys. {\bf 49,} 652 (1973).
\bibitem{seesaw}
P. Minkowski, Phys. Lett. {\bf B67,} 421 (1977).\\
T. Yanagida, in {\it Proceedings of the Workshop on
the Baryon Number of the Universe and Unified Theories}, Editors:
O. Sawada and A. Sugamoto. Tsukuba, Japan (1979). page 95.\\
P. Ramond. {\it The Family Group in Grand Unified Theories}.
CALT-68-709 (Feb 1979). Senibel Symposium. {\tt hep-ph/9809459}.\\
S.L. Glashow, NATO Adv. Study Inst. Ser. B. Phys. {\bf 59} 687 (1979).\\
M. Gell-Mann, P. Ramond and R. Slansky, in {\it Supergravity},
Editors: P. van Nieuwenhuizen and D. Freedman.
North-Holland (1979) page 315.\\
R.N. Mohapatra and G. Senjanovic, Phys. Rev. Lett. {\bf 44,} 912 (1980). 
\bibitem{textures}
P.H. Frampton and S.L. Glashow, Phys. Lett. {\bf B461,} 95 (1999).
{\tt hep-ph/9906375} 
\bibitem{marfatia}
P.H. Frampton, S.L. Glashow and D. Marfatia, Phys. Lett. {\bf 536B,} 79 (2002).
{\tt hep-ph/0201008}.
\bibitem{FGY}
P.H. Frampton, S.L. Glashow and T. Yanagida, Phys. Lett. {\bf B548,} 119 (2002).
{\tt hep-ph/0208157}.
\bibitem{Glashow}
A. Cohen and S.L. Glashow, Phys. Rev. Lett. {\bf 97,} 021601 (2006).
{\tt hep-ph/0601236};\\
{\tt hep-ph/0605036}.
\bibitem{PMNS}
B. Pontecorvo, Sov. Phys. JETP {\bf 6,} 429 (1957)
[Zh. Eksp. Teor. Fiz. {\bf 33,} 549 (1957)].\\
Z. Maki, M. Nakagawa and S. Sakata, Prog. Thoer. Phys. {\bf 28,} 870 (1962).
\bibitem{Cabibbo2}
N. Cabibbo, Phys. Lett. {\bf B72,} 333 (1978).
\bibitem{Wolfenstein}
L. Wolfenstein, Phys. Rev. {\bf D18,} 958 (1978). 
\bibitem{Ma}
E. Ma and G. Rajasekaran, Phys. Rev. {\bf D64,} 113012 (2001). {\tt hep-ph/0106291};\\
K.S. Babu, E. Ma and J.W.F. Valle, Phys. Lett. {\bf B552,} 207 (2003).
{\tt hep-ph/0206292}.
\bibitem{Ma2}
E. Ma, Mod. Phys. Lett. {\bf A20,} 2601 (2005). {\tt hep-ph/0508099};
Phys. Lett. {\bf B632,} 352 (2006). {\tt hep-ph/0508231}.\\
B. Adhikary, B. Brahmachari, A. Ghosal, E. Ma and
M.K. Parida, Phys. Lett. {\bf B638,} 345 (2006). {\tt hep-ph/0603059}.\\
E. Ma, Phys. Rev. {\bf D73,} 057304 (2006).\\
E. Ma, H. Sawanaka, and M. Tanimoto, Phys. Lett. {\bf B641,} 301 (2006).
~~~{\tt hep-ph/0606103}.\\
E. Ma, Mod. Phys. Lett. {\bf A21,} 1917 (2006).
{\tt hep-ph/0607056}.\\
E. Ma. {\tt hep-ph/0607142,}~~{\tt hep-ph/0611181,}~~{\tt hep-ph/0612013.}  
\bibitem{HPS}
P.F. Harrison, D.H. Perkins and W.G. Scott, Phys. Lett. {\bf B530, } 167 (2002).
{\tt hep-ph/0202074}.
\bibitem{Altarelli}
G. Altarelli and F. Feruglio, Nucl. Phys. {\bf B720,} 64 (2005).
{\tt hep-ph/0504165};
Nucl. Phys. {\bf B742,} 215 (2006). {\tt hep-ph/0512103};\\
G. Altarelli, F. Feruglio and Y. Lin. {\tt hep-ph/0610165};\\
G. Altarelli. {\tt hep-ph/0508053,}~~{\tt hep-ph/0610164,}~~{\tt hep-ph/0611117.}
\bibitem{expt}
B.Aharmim, {\it et al.} (SNO Collaboration), Phys. Rev. {\bf C72,} 055502 (2005).
{\tt nucl-ex/0502021}.
\bibitem{ThomasWood}
A.D. Thomas and G.V. Wood, {\it Group Tables}. Shiva Mathematical Series (1980).
\bibitem{Kephart1}
P.H. Frampton and T.W. Kephart, Int. J. Mod. Phys.  {\bf 10A,} 4689 (1995).
{\tt hep-ph/9409330}.
\bibitem{Kephart2}
P.H. Frampton and T.W. Kephart, Phys. Rev. {\bf D64,} 086007 (2001). 
{\tt hep-th/0011186}.
\bibitem{Lebed}
A. Aranda, C.R. Carone and R.F. Lebed, Phys. Lett. {\bf B474,} 170 (2000).
{\tt hep-ph/9910392}; 
Phys. Rev. {\bf D62,} 016009. {\tt hep-ph/0002044};
A. Aranda, C.R. Carone and  P. Meade, Phys . Rev. {\bf D65,} 013011 (2001). {\tt hep-ph/0109120}
\bibitem{Conference}
P.H. Frampton, {\it The Third Family is Different}. In Proceeding of 
the Fourth International Symposium on Particles, Strings and Cosmology.
Editor: K.C. Wali. World Scientific (1995).
page 63 {\tt hep-ph/9409331}; {\it Treating Top Differently from Charm and Up}. Particle Theory
and Phenomenology, Editors: K.E. Lassila, {\it et al.}
World Scientific (1996) page 229. {\tt hep-ph/9507351}.
\bibitem{Kephart3}
P.H. Frampton and T.W. Kephart, Phys. Rev. {\bf D51,} R1 (1995). {\tt hep-ph/9409324}.

\end{thebibliography}
\end{document}